\documentclass[prd,aps, preprintnumbers, showpacs,twocolumn, nofootinbib,superscriptaddress,notitlepage]{revtex4-2}
\usepackage[normalem]{ulem}
\usepackage{epsfig}
\usepackage{amsfonts}
\usepackage{amsmath}
\usepackage{slashed}
\usepackage{graphicx}
\usepackage{color}
\usepackage{mathtools}
\usepackage{subfigure}
\usepackage{graphicx}

\usepackage{stmaryrd}
\usepackage{amssymb,psfrag}
\usepackage[normalem]{ulem}
\usepackage{caption}

\newcommand{\beq}{\begin{eqnarray}}
\newcommand{\eeq}{\end{eqnarray}}

\newcommand{\nn}{\nonumber}

\begin{document}

\title{Nuclear effects in extracting $\bf{\sin^2\theta_W}$ and a probe for short-range correlations}

\author{Xing-Hua Yang}
\email{yangxinghua@sdut.edu.cn}
\affiliation{School of Physics and Optoelectronic Engineering, Shandong University of Technology, Zibo, Shandong 255000, China}

\author{Fei Huang}
\email{Corresponding author. fhuang@sjtu.edu.cn}
\affiliation{School of Physics and Technology, University of Jinan, Jinan, Shandong 250022, China}
\affiliation{INPAC, Key Laboratory for Particle Astrophysics and Cosmology (MOE),
Shanghai Key Laboratory for Particle Physics and Cosmology,
School of Physics and Astronomy, Shanghai Jiao Tong University, Shanghai 200240, China}

\author{Ji Xu}
\email{Corresponding author. xuji\_phy@zzu.edu.cn}
\affiliation{School of Physics and Microelectronics, Zhengzhou University, Zhengzhou, Henan 450001, China}

\begin{abstract}
  We investigate the neutral-current neutrino-nucleon deep inelastic scattering with particular emphasis on short-range correlation and EMC effect, as well as their impact on the weak-mixing angle $\sin^2\theta_W$ determination. The ratios of structure function $F_{2(NC)}^{A}$ and $x F_{3(NC)}^{A}$ are presented where the nuclei $A$ are chosen as carbon, iron and lead. One kind of universal modification function is proposed which would provide a nontrivial test of SRC universality on the platform of neutrino-nucleon DIS. In addition, we study the impact of ``SRC-driven'' nuclear effects on the extraction of $\sin^2\theta_W$ which is naturally associated with the renowned NuTeV anomaly. The results indicate that these effects may account for a substantial fraction of the NuTeV anomaly and considerably affect the value of extracted $\sin^2\theta_W$.
\end{abstract}

\maketitle

\section{Introduction}
\label{sec:Introduction}
Neutrino-nucleon scattering provides one of the most precise platforms for the weak neutral current. The high statistics measurements of neutrino deep inelastic scattering (DIS) on heavy nuclear targets have attracted lots of attention due to their importance in global fits
of parton distribution functions (PDFs). Besides, the data on $\nu$ and $\bar \nu$, contrary to charged leptons, give direct access to both the weak-mixing angle $\theta_W$ and the $Z^0$ coupling to (anti)neutrinos.

Owing to the weak nature of neutrino interactions, the use of heavy nuclear targets is unavoidable in neutrino DIS experiments, and this complicates the extraction of relevant observables because of the nuclear effects. The original idea of having nuclear effects in PDFs was driven by data in DIS measurements performed by the European Muon Collaboration (EMC) and subsequently conformed by other experiments \cite{Aubert:1983xm,Arneodo:1988aa,Arneodo:1989sy,Allasia:1990nt,Gomez:1993ri,Seely:2009gt}. They found a reduction of the cross section per nucleon in nucleus $A$ compared with deuteron in the valence quark dominated regime $0.3<x<0.7$, here $x$ is the Bjorken variable. This phenomenon is referred to as the EMC effect. One should note that there is no consensus on the exact nature of EMC effect until now.

Recently, the possible connection between the EMC effect and short-range correlation (SRC) has been investigated substantially \cite{Weinstein:2010rt,Egiyan:2005hs,Hen:2014nza,Duer:2018sby,Chen:2016bde,Xu:2019wso,Hatta:2019ocp,Lynn:2019vwp,Huang:2021cac,Hu:2021fxa,Bertone:2018dse,Wang:2022kwg}. The two-nucleon SRC is defined experimentally as having small center-of-mass momentum and large relative momentum, it describes the probability that two nucleons are close in coordinate space, as a result of nontrivial nucleon-nucleon interactions in the nucleus. One can refer to this nicely written review \cite{Hen:2016kwk} for more details. The neutrino-nucleon DIS is an ideal platform for testing nucleon structures and SRC interpretation of the EMC effect, in this work, we will study the neutral-current neutrino-nucleon DIS where the nuclei $A$ are selected to be $\rm{^{12}C}$, $\rm{^{56}Fe}$ and $\rm{^{208}Pb}$. The structure functions $F_{2(NC)}^{A}(x,Q^2)$ and $x F_{3(NC)}^{A}(x,Q^2)$ are calculated with consideration of nuclear PDFs (NPDFs) in terms of EPPS21 parametrization, and we choose CT18ANLO as free nucleon baseline \cite{Eskola:2021nhw,Hou:2019efy}. One kind of universal modification functions was proposed in this process which can be viewed as nontrivial tests of SRC universality on the platform of neutrino-nucleon scattering. Here universality means the partonic structure from the correlated nucleon-nucleon SRC pair is the same for all kinds of nuclei.

Compared to charged lepton-nucleon scattering, data on neutrino-nucleon are in short supply, and the understanding of nuclear effects in them are therefore pretty limited. In addition to testing the SRC universality, there is an even more important issue, we explore the way in which the ``SRC-driven'' nuclear effects modify the extraction of weak-mixing angle $\sin^2\theta_W$, which is a key parameter in the electroweak sector of the Standard Model (SM) \cite{Kumar:2013yoa,Marquet:2019ltn,Wei:2020glg}. The precise determination of this angle is among the fundamental works in particle physics and it had been accurately measured by collider experiments. Nevertheless, the NuTeV Collaboration reported an anomalously large weak-mixing angle $\sin^2\theta_W = 0.2277 \pm 0.0013 \,(\textrm{stat}) \pm 0.0009 \,(\textrm{syst})$ \cite{NuTeV:2001whx,NuTeV:2002ryj}. There is a three-sigma discrepancy between this result and global analysis of other data $\sin^2\theta_W = 0.2227 \pm 0.0004$. This is the renowned NuTeV anomaly, which has not been fully understood yet. Since this anomaly came up, a lot of detailed data analyzing works followed, the results raise a deep question as to whether the neutrino DIS data could be combined with the charged-lepton DIS data to get better NPDFs \cite{Tzanov:2005kr,Schienbein:2007fs,Schienbein:2009kk,Hirai:2009mq,Kovarik:2010uv,Paukkunen:2010hb}.

This situation requires resolution. Historically, the precise measurement of $\theta_W$ is closely related to new physics (NP) \cite{Georgi:1974sy,Georgi:1974yf,Marciano:1990dp}, and many mechanisms based on NP were proposed to explain the cause of this anomaly \cite{Davidson:2001ji,Kurylov:2003by}. Meanwhile, a number of works that attempted to interpret the NuTeV result within the context of the SM have been suggested, and most of them have the potential to attenuate the anomaly \cite{Londergan:2003ij,Martin:2003sk,Kovalenko:2002xe,Kumano:2002ra,Gluck:2005xh,Kulagin:2007ju,Ball:2009mk,Cloet:2009qs,Yang:2022xwy,Bentz:2009yy,Londergan:1998ai}. These works largely focused on nucleon charge symmetry violating (CSV) effects, finite distributions of strange sea quarks, as well as nuclear corrections such as Fermi motion and binding and the isovector EMC effect. If one or more contributions mentioned above are as large as expected in the references, it undoubtedly will be a milestone discovery concerning fundamental QCD effects in nuclei. In spite of these remarkable efforts, effects from the SRCs of the bound nucleon have not been investigated in relation to the NuTeV anomaly. These effects are potentially essential since they are widely accepted as one of the leading approaches for explaining the EMC effect.

It is known that the Paschos-Wolfenstein (PW) relation which was deduced for the isoscalar nucleon $R_A^{-}= \left(\sigma_{N C}^{\nu A}-\sigma_{N C}^{\bar{\nu} A}\right) / \left(\sigma_{C C}^{\nu A}-\sigma_{C C}^{\bar{\nu} A}\right) = 1 / 2-\sin ^2 \theta_W$ \cite{Paschos:1972kj}, was used for the determination of $\sin^2\theta_W$ in NuTeV experiment. Here, $\sigma_{N C}^{\nu A}$ and $\sigma_{C C}^{\nu A}$ are the deep inelastic cross sections for neutral-current (NC) and charged-current (CC) neutrino-nucleon interactions, and $A$ represents the target. In this paper, motivated by the correlation between the EMC effect and the SRC scale factor, we derive a modified PW relation. Then, we discuss its impact on the extraction of $\sin^2\theta_W$.

The rest of this paper is organized as follows: In Sec.\,\ref{Universality}, the formalism and results of structure functions in NC neutrino-nucleon DIS will be briefly reviewed, and the proposal of one kind of universal modification function is discussed. In Sec.\,\ref{weakangle} we study the ``SRC-driven'' nuclear corrections of the PW relation and their possible effects on the extraction of $\sin^2\theta_W$. We conclude in Sec.\,\ref{summary}.

\section{Universal modification function in NC neutrino-nucleon DIS}
\label{Universality}
\begin{figure}
\includegraphics[width=0.58\columnwidth]{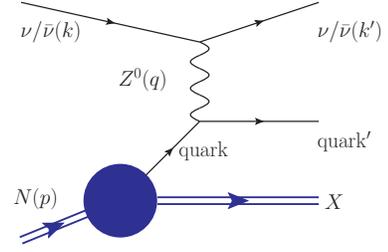}
\caption{A schematic diagram for neutral-current neutrino-nucleon DIS. The process was carried out through the exchange of an electroweak gauge boson $Z^0$.}
\label{NC_DIS}
\end{figure}

As illustrated in Fig.\,\ref{NC_DIS}, a high energy neutrino interacts with a nucleon through the exchange of a neutral $Z^0$ boson, producing a corresponding neutrino and hadron in the final states. These processes can be analyzed by the following Lorentz invariants: the Bjorken scaling variable $x\equiv Q^2/(2p\cdot q)$; the inelasticity $y\equiv(2p\cdot k)/(2p\cdot q)$; the negative squared four momentum transfer $Q^2\equiv -q^2$.

The cross section for NC (anti)neutrino interactions with the nucleon in nucleus $A$ is given by
\begin{eqnarray}\label{diff_for_nuandbarnu1}
  \frac{d \sigma_{N C}^{\nu A}}{d x d y} &=& \frac{G_F^2 s}{2 \pi\left(1+Q^2 / M_Z^2\right)^2}\left[F_{1(N C)}^{\nu A} x y^2 \right. \nn\\
  && \left. +F_{2(N C)}^{\nu A}\left(1-y\right) + F_{3(N C)}^{\nu A} x y\left(1-\frac{y}{2}\right)\right] \,,\nn\\
  \frac{d \sigma_{N C}^{\bar\nu A}}{d x d y} &=& \frac{G_F^2 s}{2 \pi\left(1+Q^2 / M_Z^2\right)^2}\left[F_{1(N C)}^{\bar\nu A} x y^2 \right. \nn\\
  && \left. +F_{2(N C)}^{\bar\nu A}\left(1-y\right) - F_{3(N C)}^{\bar\nu A} x y\left(1-\frac{y}{2}\right)\right] \,.
\end{eqnarray}
Here $G_F$ is the Fermi coupling constant and $s$ is the square of the center-of-mass energy. The expressions above can be reduced by Callan-Gross relation $F_{2(NC)}^{\nu A}=2x F_{1(NC)}^{\nu A}, F_{2(NC)}^{\bar\nu A}=2x F_{1(NC)}^{\bar\nu A}$ \cite{Callan:1969uq}. In the above expression, we have omitted the explicit dependence on $x$ and $Q^2$ for brevity. The structure functions for a nucleus with the mass number $A$ and the atomic number $Z$ can be decomposed into two parts
\begin{eqnarray}
  F_{2(N C)}^{\nu A} &=& \frac{Z}{A}F_{2(N C)}^{\nu p/ A} + \frac{A-Z}{A}F_{2(N C)}^{\nu n/ A} \,,\nn\\
  F_{2(N C)}^{\bar\nu A} &=& \frac{Z}{A}F_{2(N C)}^{\bar\nu p/ A} + \frac{A-Z}{A}F_{2(N C)}^{\bar\nu n/ A} \,,
\end{eqnarray}
which can be further expressed in terms of NPDFs:
\begin{eqnarray}\label{calcstrucuturefuncs1}
  F_{2(N C)}^{A} &=& F_{2(N C)}^{\nu A} + F_{2(N C)}^{\bar\nu A} \nn\\
  &=& 4x\frac{Z}{A} \Big[ (u_L^2 +u_R^2)(u_p^{A+} + c_p^{A+}) \nn\\
  &&+(d_L^2+d_R^2)(d_p^{A+} +s_p^{A+}) \Big] \nn\\
  &&+ 4x\frac{A-Z}{A} \Big[ (u_L^2 +u_R^2)(u_n^{A+} + c_n^{A+}) \nn\\
  &&+(d_L^2+d_R^2)(d_n^{A+} +s_n^{A+}) \Big] \,.
\end{eqnarray}
Here the left- and right-handed couplings for a quark are expressed as $u_L=\frac{1}{2}-\frac{2}{3}\sin^2\theta_W, u_R=-\frac{2}{3}\sin^2\theta_W$ and $d_L=-\frac{1}{2}+\frac{1}{3}\sin^2\theta_W, d_R=\frac{1}{3}\sin^2\theta_W$. We also define $q^{A\pm}\equiv q^{A} \pm \bar q^{A}$.

The size of partonic CSV correction to NuTeV anomaly has been estimated to be significant \cite{Bentz:2009yy,Londergan:1998ai}. Nevertheless, since this correction has been relatively well studied and we are only interested in the SRC-induced nuclear corrections, the isospin symmetry $u_p^A(\bar u_p^A) = d_n^A(\bar d_n^A) \,, d_p^A(\bar d_p^A) = u_n^A(\bar u_n^A) \,, s_p^A(\bar s_p^A) = s_n^A(\bar s_n^A) \,, c_p^A(\bar c_p^A) = c_n^A(\bar c_n^A)$ will be utilized in this paper. Therefore, the expression for the structure function in Eq.\,(\ref{calcstrucuturefuncs1}) simplifies to
\begin{eqnarray}\label{calcstrucuturefuncs2}
  F_{2(NC)}^{A} &=& 2\frac{Z}{A}2x \Big[ (u_L^2 +u_R^2)u_p^{A+} +(d_L^2+d_R^2)d_p^{A+} \Big] \nn\\
  &&+ 2\frac{A-Z}{A}2x \Big[ (u_L^2 +u_R^2)d_p^{A+} +(d_L^2+d_R^2)u_p^{A+} \Big] \nn\\
  &&+ 4x \Big[ (u_L^2 +u_R^2)(c_p^{A+}) +(d_L^2+d_R^2)s_p^{A+} \Big] \,.
\end{eqnarray}
Similarly, the $x F_{3(NC)}^{A}$ is
\begin{eqnarray}\label{alcstrucuturefuncs3}
  x F_{3(NC)}^{A} &=& 2\frac{Z}{A}2x \Big[ (u_L^2-u_R^2)u_p^{A-} +(d_L^2-d_R^2)d_p^{A-} \Big] \nn\\
  &&+ 2\frac{A-Z}{A}2x \Big[ (u_L^2-u_R^2)d_p^{A-} +(d_L^2-d_R^2)u_p^{A-} \Big] \nn\\
  &&+ 4x \Big[ (u_L^2-u_R^2)c_p^{A-} + (d_L^2-d_R^2)s_p^{A-} \Big] \,.
\end{eqnarray}
The NPDF $f_i^{p/A}(x,Q^2)$ can be defined relative to the free proton PDF $f_i^{p}(x,Q^2)$ as \cite{Eskola:2021nhw}
\begin{eqnarray}\label{EPPS21R}
  f_{i}^{p / A}\left(x, Q^{2}\right)=R_{i}^{A}\left(x, Q^{2}\right) f_{i}^{p}\left(x, Q^{2}\right) \,,
\end{eqnarray}
where $i$ denotes the types of partons and $R_{i}^{A}\left(x, Q^{2}\right)$ refers to nuclear modification factor. The free proton baseline is CT18ANLO \cite{Hou:2019efy}.

\begin{figure}[htbp]
\centering
\includegraphics[width=0.78\columnwidth]{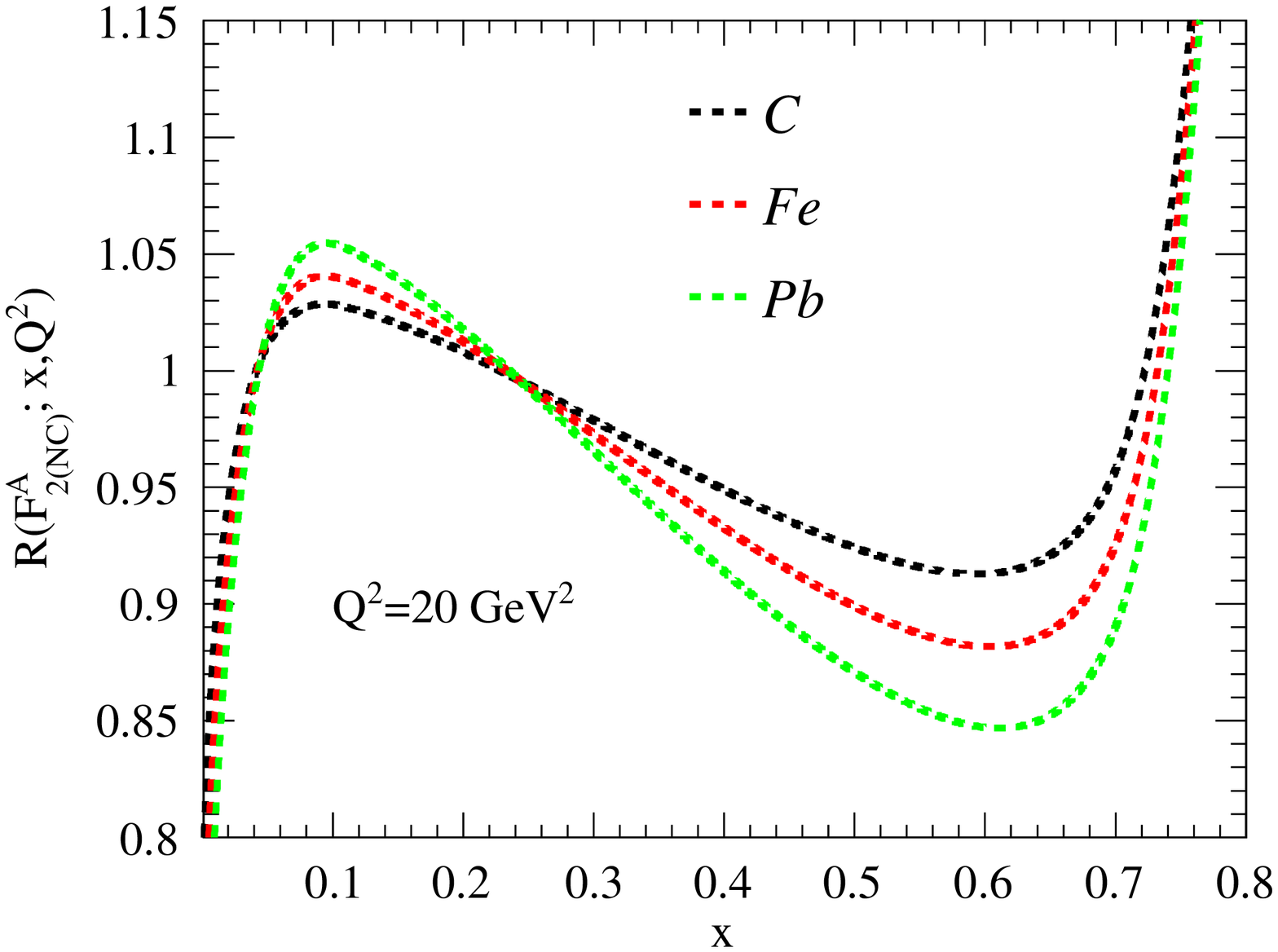}
\centering
\includegraphics[width=0.78\columnwidth]{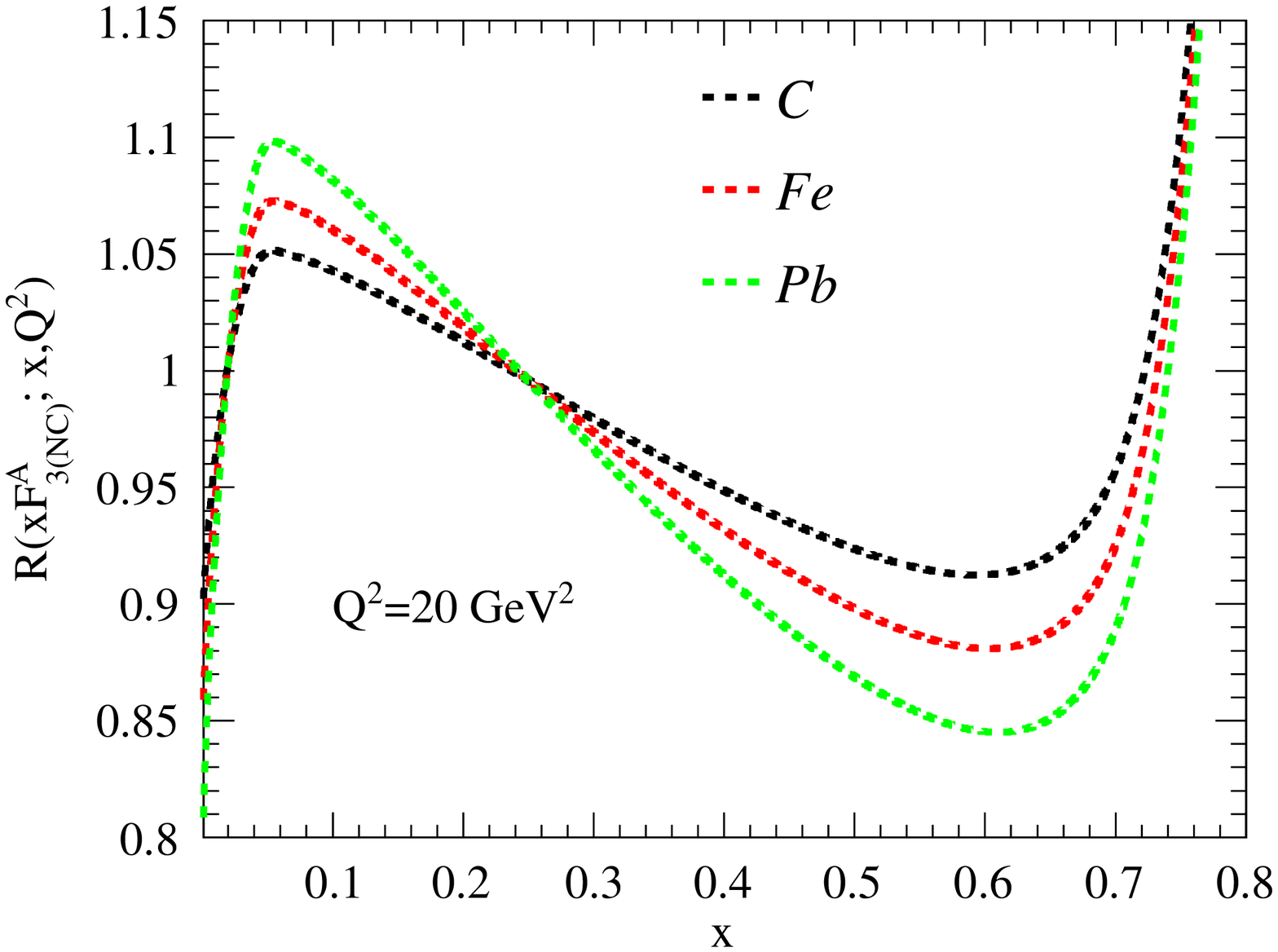}
\centering
\caption{The ratios \textbf{(a)} $R(F_{2(NC)}^A;x,Q^2)$ and \textbf{(b)} $R(xF_{3(NC)}^A;x,Q^2)$ as functions of $x$ with $Q^2=20\,\textrm{GeV}^2$. The black, red and green lines correspond to $\rm{C}$, $\rm{Fe}$ and $\rm{Pb}$ respectively.}
\label{R_structure_functions}
\end{figure}

\begin{figure}[htbp]
\centering
\includegraphics[width=0.78\columnwidth]{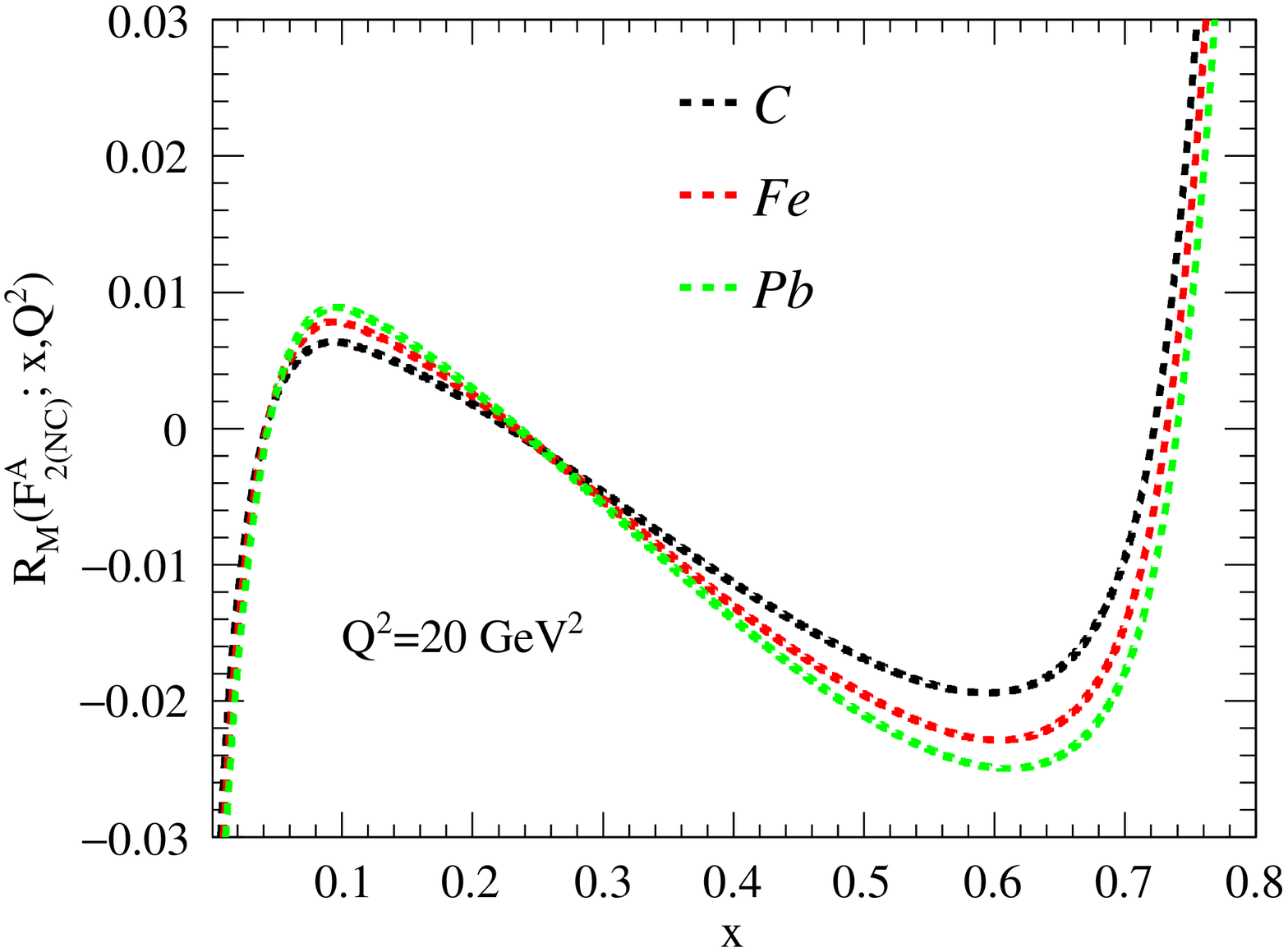}
\centering
\includegraphics[width=0.78\columnwidth]{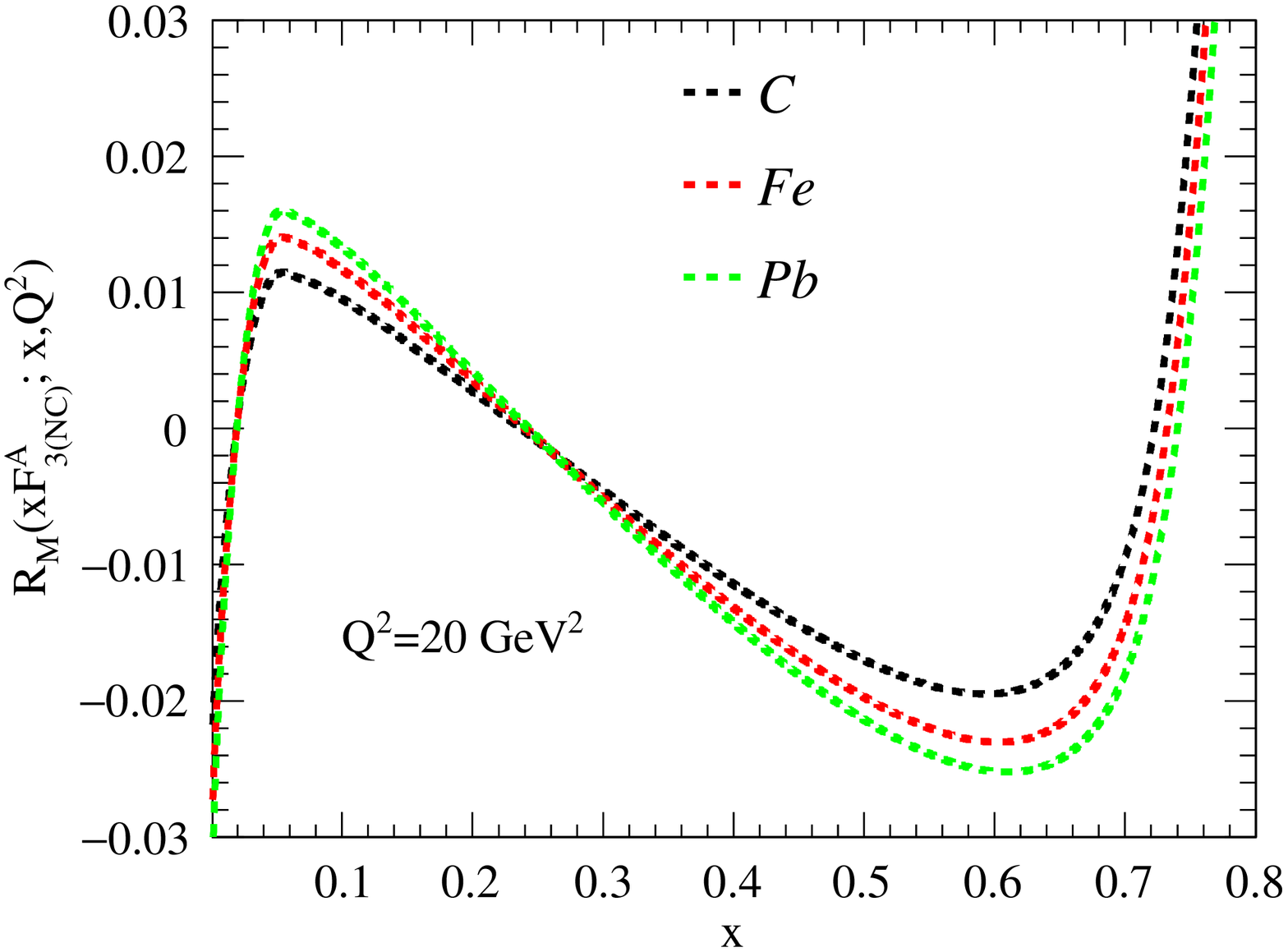}
\centering
\caption{The ratios \textbf{(a)} $R_M(F_{2(NC)}^A;x,Q^2)$ and \textbf{(b)} $R_M(xF_{3(NC)}^A;x,Q^2)$ present the universality of SRC contributions with EPPS21 parametrization.}
\label{Ex_EPPS21}
\end{figure}

We define the ratios in line with Ref.\,\cite{Huang:2021cac}
\begin{eqnarray}\label{RF2andRF3}
  R(F_{2(NC)}^{A};x,Q^2) &=& F_{2(NC)}^{A}/F_{2(NC)} \,,\nn\\
  R(x F_{3(NC)}^{A};x,Q^2) &=& \big( x F_{3(NC)}^{A} \big) / \big( x F_{3(NC)} \big) \,.
\end{eqnarray}
Here $F_{2(NC)}$ and $x F_{3(NC)}$ have the same expressions of $F_{2(NC)}^{A}$ and $xF_{3(NC)}^{A}$, just with the NPDFs in nucleon replaced by PDFs in free proton. With Eq.\,(\ref{EPPS21R}) and Eq.\,(\ref{RF2andRF3}), we depict the dependence of $R(F_{2(NC)}^{A};x,Q^2)$ and $R(x F_{3(NC)}^{A};x,Q^2)$ on $x$ in Fig.\,\ref{R_structure_functions}. The $Q^2$ is fixed to $20\,\textrm{GeV}^2$, which is attainable in many neutrino-nucleon scattering experiments. A comparison will be conducted between this figure and Fig.\,\ref{Ex_EPPS21} that appears subsequently.

Motivated by the amazing linear correlation between the EMC effect and the SRC scale factor which has received enormous attention in recent years, we parameterize the $u$ and $d$ quark distributions in the EMC region as that for the structure function in Ref.\,\cite{Frankfurt:1993sp,Segarra:2019gbp} by assuming that all nuclear modifications originate from the nucleon-nucleon SRCs,
\begin{eqnarray}\label{para_about_distributions}
  u_v^{p/A}\left(x, Q^{2}\right) \!&=&\! \frac{1}{Z}\! \left[Z u_v^{p}\left(x, Q^{2}\right)+n_{s r c}^{A} \, \delta u_v^{p}\left(x, Q^{2}\right)\right] \,, \nn\\
  d_v^{p/A}\left(x, Q^{2}\right) \!&=&\! \frac{1}{Z}\! \left[Z d_v^{p}\left(x, Q^{2}\right)+n_{s r c}^{A} \, \delta d_v^{p}\left(x, Q^{2}\right)\right] \,,
\end{eqnarray}
where $n^A_{src}$ represents number of nucleon-nucleon pairs in nucleus $A$, notably the subscript $v$ in $u_v^{p}$ and $d_v^{p}$ means distributions of valence quark since experimental results pointed to the EMC effect being due to a change in the valence quark distributions \cite{Alde:1990im,Bertsch:1993vx}. $\delta u_v^{p}$ and $\delta d_v^{p}$ represent the difference between $u$ and $d$ valence quark distributions in the SRC pair and in the free proton respectively.

We then rearrange equation Eq.\,(\ref{para_about_distributions}) with the aid of parametrization in Eq.\,(\ref{EPPS21R}):
\begin{eqnarray}\label{uni_function_of_ud}
  \delta u_v^p\left(x, Q^{2}\right) / u_v^p\left(x, Q^{2} \right) &=& \left( R_{u_v}^{A}\left(x, Q^{2}\right)-1 \right) / \left( n_{src}^A/Z_A \right) \,,\nn\\
  \delta d_v^p\left(x, Q^{2}\right) / d_v^p\left(x, Q^{2}\right) &=& \left( R_{d_v}^{A}\left(x, Q^{2}\right)-1\right) / \left(n_{src}^A/Z_A \right) \,.\nn\\
\end{eqnarray}
Because $\delta u_v^p$ and $\delta d_v^p$ are assumed to be nucleus-independent, our model predicts that the left-hand side of Eq.\,(\ref{uni_function_of_ud}) should be a universal function, here universal means they are the same for all kinds of nuclei. This indicates that the nucleus-dependent quantities on the right-hand side of Eq.\,(\ref{uni_function_of_ud}) combine to give a nucleus-independent result. This universality of SRC can be illustrated more specifically by introducing one kind of universal modification functions
\begin{eqnarray}
  \!\!\!\!\!\!\!\! R_{M}(F_{2(NC)}^{A};x,Q^2) \!&=&\! \frac{2Z_A}{A_A} \! \frac{R(F_{2(NC)}^{A} ;x,Q^2)-1}{a_2^{A}} \,,\nn\\
  \!\!\!\!\!\!\!\! R_{M}(x F_{3(NC)}^{A};x,Q^2) \!&=&\! \frac{2Z_A}{A_A} \! \frac{R(x F_{3(NC)}^{A} ;x,Q^2)-1}{a_2^{A}} \,.
\end{eqnarray}
Here $a_{2}^{A}=\left(n_{s r c}^{A} / A\right) /\left(n_{s r c}^{d} / 2\right)$ is the SRC scale factor of nucleus $A$ respect to that of deuteron which can be measured through the nuclear structure functions at $x\!>1.5\!$ region \cite{Fomin:2011ng,Arrington:2012ax}. In fact, the role of $R_{M}$ is normalizing the ratios defined in Eq.\,(\ref{RF2andRF3}) by the respective SRC scale factors. The universal modification functions are plotted in Fig.\,\ref{Ex_EPPS21}, it can be seen that the ratios of different nuclei tend to shrink substantially compared with those in Fig.\,\ref{R_structure_functions}. This universality has been investigated in the CC neutrino-nucleon DIS previously \cite{Huang:2021cac}, our result in the NC process here supports the theoretical assumption raised in Eq.\,(\ref{para_about_distributions}) and indicates that in the valence quark dominated regime $0.3\!<\!x\!<\!0.7$, the EMC effect is mainly caused by SRC pairs. Fig.\,\ref{Ex_EPPS21} is a nontrivial test of SRC universality, providing a new clue to understand how the relatively long-range nuclear interaction influences the short-distance parton structure inside the nucleon.

\section{Modified PW relation and its impact on $\bf{\sin^2\theta_W}$ determination}
\label{weakangle}
The NuTeV target is mainly the iron nuclei, nuclear corrections should be carefully taken into account for a precise determination of $\sin^2\theta_W$. In this section, we derive a modified PW relation that has taken the SRC of nucleons into account. Then, we discuss a possible nuclear modification factor that could change the extracted $\sin^2\theta_W$ value.

The differential cross sections of charged-current (anti)neutrino-nucleon DIS expressed in terms of NPDFs are:
\begin{eqnarray}\label{CC_diff}
  \frac{d \sigma_{C C}^{\nu A}}{d x d y} &=& \frac{G_F^2 s}{\pi\left(1+Q^2 / M_W^2\right)^2} x \nn\\
   &&\times\left[ d^A +s^A +(1-y)^2(\bar u^A +\bar c^A) \right] \,,\nn\\
  \frac{d \sigma_{C C}^{\bar\nu A}}{d x d y} &=& \frac{G_F^2 s}{\pi\left(1+Q^2 / M_W^2\right)^2} x \nn\\
   &&\times\left[ \bar d^A +\bar s^A +(1-y)^2(u^A + c^A) \right] \,.
\end{eqnarray}
For the NC neutrino-nucleon scattering, the differential cross sections are:
\begin{eqnarray}\label{NC_diff}
\frac{d \sigma_{N C}^{\nu A}}{d x d y} &=& \frac{G_F^2 s}{\pi\left(1+Q^2 / M_Z^2\right)^2} x \bigg\{ [u_L^2 +u_R^2(1-y)^2](u^A+c^A) \nn\\
   && +[u_R^2 +u_L^2(1-y)^2](\bar u^A +\bar c^A) \nn\\
   &&+[d_L^2 +d_R^2(1-y)^2](d^A +s^A) \nn\\
   &&+[d_R^2 +d_L^2(1-y)^2](\bar d^A +\bar s^A) \bigg\} \,, \nn\\
\frac{d \sigma_{N C}^{\bar\nu A}}{d x d y} &=& \frac{G_F^2 s}{\pi\left(1+Q^2 / M_Z^2\right)^2} x \bigg\{ [u_R^2 +u_L^2(1-y)^2](u^A + c^A) \nn\\
   &&+[u_L^2 +u_R^2(1-y)^2](\bar u^A + \bar c^A) \nn\\
   &&+[d_R^2 +d_L^2(1-y)^2]( d^A + s^A ) \nn\\
   &&+[d_L^2 +d_R^2(1-y)^2](\bar d^A + \bar s^A) \bigg\} \,.
\end{eqnarray}
Utilizing Eq.\,(\ref{CC_diff}) and Eq.\,(\ref{NC_diff}), we obtain the PW relation in the form of parton distributions \cite{Kumano:2002ra},
\begin{eqnarray}\label{expreofAR-}
  \!\!\!\! R_A^- \!&=&\! \left(1-(1-y)^2\right) \nn\\
  &\!\times&\! \frac{ (u_L^2-u_R^2)(u_v^A +c_v^A) +(d_L^2-d_R^2)(d_v^A+s_v^A)}{ d_v^A + s_v^A -(1-y)^2(u_v^A +c_v^A)} \,.
\end{eqnarray}
The valence quark PDFs are defined by $q_v^A \equiv q^A -\bar q^A$. We can parameterize $u_v^A$ and $d_v^A$ in the EMC region as we did in Eq.\,(\ref{para_about_distributions}), therefore for a certain nucleus with the mass number $A$, the atomic number $Z$ and neutron number $N$, it's valence $u$ and $d$ quark distributions would be decomposed into contributions from unmodified protons and neutrons and $np$ SRC pairs with modified quark distributions:
\begin{eqnarray}\label{uva1}
  u_v^A &=& \frac{1}{A}\left[ (Z-n_{src}^A)u_v^p + (N-n_{src}^A)u_v^n + n_{src}^A\left(\tilde u_v^p +\tilde u_v^n\right) \right] \nn\\
  &=& \frac{1}{A}\left[ Z u_v^p +N d_v^p +n_{src}^A(\delta u_v^p+\delta u_v^n) \right] \,.
\end{eqnarray}
Here the partonic charge symmetry between free proton and neutron $u_v^n = d_v^p$ has been utilized. $\tilde u_v^p$ and $\tilde u_v^n$ are the modified distributions for protons and neutrons in SRC pairs and $\delta u_v^p \equiv \tilde u_v^p - u_v^p$ (similarly for $\delta u_v^n$). It is analogous for $d_v^A$,
\begin{eqnarray}\label{dva1}
   d_v^A = \frac{1}{A}\left[ Z d_v^p +N u_v^p +n_{src}^A(\delta d_v^p+\delta d_v^n) \right] \,.
\end{eqnarray}
In nucleus $A$, the isospin symmetry $u_v^{p/A}=d_v^{n/A}$ and $d_v^{p/A}=u_v^{n/A}$ restricts:
\begin{eqnarray}\label{relation_between_delta}
 \delta d_v^n = \frac{N}{Z} \delta u_v^p \,,\qquad \delta u_v^n = \frac{N}{Z} \delta d_v^p \,.
\end{eqnarray}
This makes Eq.\,(\ref{uva1}) and Eq.\,(\ref{dva1}) into
\begin{eqnarray}\label{uva2_and_dva2}
  u_v^A &=& \frac{Z u_v^p +N d_v^p}{A} + \frac{n_{src}^A}{A} \left( \delta u_v^p +\frac{N}{Z}\delta d_v^p \right) \,,\nn\\
  d_v^A &=& \frac{Z d_v^p +N u_v^p}{A} + \frac{n_{src}^A}{A} \left( \delta d_v^p +\frac{N}{Z}\delta u_v^p \right) \,.
\end{eqnarray}

Next, we define neutron excess constant $\epsilon_n$ as well as $\Delta^+$, $\Delta^-$ by
\begin{eqnarray}\label{sub3}
  && \epsilon_n \equiv \frac{N-Z}{A} \,,\quad \Delta^+ \equiv \delta d^p_v + \delta u^p_v  \,,\quad  \Delta^- \equiv \delta d^p_v - \delta u^p_v  \,. \nn\\
\end{eqnarray}
Substituting Eqs.\,(\ref{uva2_and_dva2}) and Eq.\,(\ref{sub3}) together with the coupling constants into Eq.\,(\ref{expreofAR-}), we can reexpress $R_A^-$:
\begin{eqnarray}\label{full_RA-}
  &&R_A^- \!=\! \bigg[y(y-2) \Big( A(1+\epsilon_n)\big[ s_v^A-c_v^A \nn\\
  &&+(3u_v^p+3d_v^p+2s_v^A+4c_v^A)\cos\!2\theta_W \!+\! 2\epsilon_n(d_v^p-u_v^p) \sin^2\!\theta_W \big] \nn\\
  &&+n_{src}^A \big[ 4\epsilon_n\Delta^- \sin^2\!\theta_W +6\Delta^+ \cos\!2\theta_W \big] \Big) \bigg] \bigg/ \bigg[ 6\Big( A(1+\epsilon_n) \nn\\
  &&\times \big[ y(y-2)(u_v^p+d_v^p)-(2-2y+y^2)\epsilon_n(d_v^p-u_v^p) \nn\\
  &&-2s_v^A+2(1-y)^2c_v^A \big] +2n_{src}^A \big[(-2+2y-y^2)\epsilon_n \Delta^- \nn\\
  &&+y(y-2)\Delta^+ \big] \Big)\bigg] \,.
\end{eqnarray}
The neutron excess effects, i.e., the $\epsilon_n$ terms, have been taken into account in the NuTeV analysis \cite{NuTeV:2001whx}, with the assumption that the target is composed of free nucleons. The ``valence'' strange and charm distributions are very tiny, if not negligible \cite{NuTeV:2002ryj,Bentz:2009yy}. These sources of corrections would not be discussed in this work, since we are interested in finding out the influence of the ``SRC-driven'' nuclear effects, i.e., $\Delta^\pm$ related terms. We reduce $R_A^-$ in Eq.\,(\ref{full_RA-}) by considering that these terms are small, retain only the leading power corrections,
\begin{eqnarray}\label{RA-final}
  R_A^- &=& \frac{1}{2}-\sin^2\theta_W \nn\\
  &&-(s_v^A -c_v^A)\frac{n_{src}^A}{A} \left(\Delta^+\right) \nn\\
  &&\times\frac{y(y-2)\big( y(y-2)+2(3+y(y-2))\cos2\theta_W \big)}{3\big(y(y-2)(u_v^p+d_v^p)\big)^2} \nn\\
  &&+\epsilon_n \frac{n_{src}^A}{A} \left(\Delta^-\right) \nn\\
  &&\times\frac{y(y-2)+2(3+y(y-2))\cos2\theta_W}{3y(y-2) \big( u_v^p+d_v^p+2\frac{n_{src}^A}{A}\Delta^+ \big) } \nn\\
  && +\mathcal{O}(\textrm{other corrections}) \,.
\end{eqnarray}
The first term is the PW relation \cite{Paschos:1972kj}. The second and third terms are corrections caused by $\Delta^{\pm}$, here ``other corrections'' means the higher corrections of $\Delta^{\pm}$ and corrections of $\epsilon_n, s_v^A, c_v^A$ which are not related to $\Delta^{\pm}$. We note reader that the $u_v^p$ and $d_v^p$ are PDFs of free proton.

In order to explore whether the $\Delta^{\pm}$ corrections could explain, or at least partially, the NuTeV anomaly, we combine Eq.\,(\ref{EPPS21R}) and Eq.\,(\ref{para_about_distributions}) to obtain estimates of $n_{src}^A \Delta^{\pm}$,
\begin{eqnarray}
  n_{src}^A(\Delta^+) = Z \left[ (R_{d_v}^A-1) d_v^p + (R_{u_v}^A-1) u_v^p \right] \,,\nn\\
  n_{src}^A(\Delta^-) = Z \left[ (R_{d_v}^A-1) d_v^p - (R_{u_v}^A-1) u_v^p \right] \,.
\end{eqnarray}
The contribution of the second term in Eq.\,(\ref{RA-final}) is almost zero since it is proportional to $(s_v^A -c_v^A)$ and the contribution of the third term is related to neutron excess constant.

\begin{figure}[htbp]
\centering
\includegraphics[width=0.78\columnwidth]{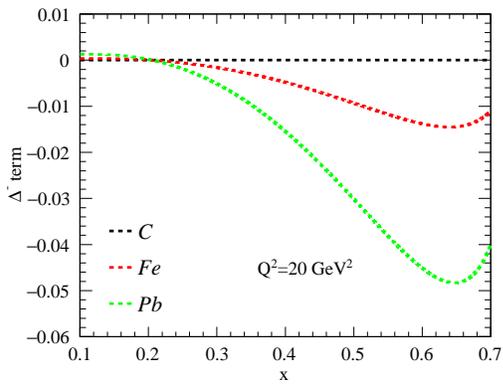}
\centering
\caption{The $\Delta^-$ correction term is evaluated at $Q^2= 20\,\textrm{GeV}^2$. The black, red and green lines correspond to $\rm{C}$, $\rm{Fe}$ and $\rm{Pb}$ respectively.}
\label{Delta-}
\end{figure}

In the NuTeV measurements, $97\%$ of the data is contained within $1\,\textrm{GeV}^2\!<\!Q^2\!<\!140\,\textrm{GeV}^2$, $0.01\!<\!x\!<\!0.75$. The average $Q^2 \!=\! 25.6\,\textrm{GeV}^2$ and $E_\nu \!=\! 120\,\textrm{GeV}$ for $\nu$ events as well as $Q^2 \!=\! 15.4\,\textrm{GeV}^2$ and $E_{\bar\nu} \!=\! 112\,\textrm{GeV}$ for $\bar \nu$ events. In our estimation, we have adopted $Q^2 \!=\! 20\,\textrm{GeV}^2$, $E \!=\! 116\,\textrm{GeV}$. Fig.\,\ref{Delta-} shows the shape of $\Delta^-$ correction term as a function of $x$. As one can see, the magnitude of this correction is strongly dependent on the neutron excess constant which is $0/12$, $4/56$ and $44/208$ for $\rm{^{12}C}$, $\rm{^{56}Fe}$ and $\rm{^{208}Pb}$, respectively. Another essential feature illustrated in this figure is that the functions change their signs at $x\approx 0.25$ where the transition from antishadowing region to the EMC region takes place. The $\Delta^-$ correction term of $\rm{^{56}Fe}$ implies a resulting decrease in the NuTeV value of $\sin^2\theta_W$ and becomes comparable to the NuTeV deviation $(0.2227-0.2277=-0.0050)$.

\begin{figure}[htbp]
\centering
\includegraphics[width=0.78\columnwidth]{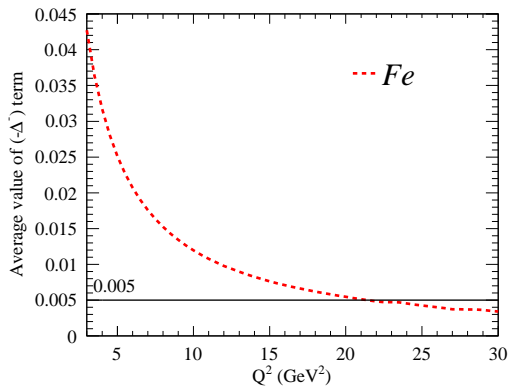}
\centering
\caption{The red dashed line indicates the negative value of $\Delta^-$ term averaged on $x$. The black solid line refers to the NuTeV deviation for convenience of comparison.}
\label{Averageonx}
\end{figure}

We then proceed to simply average the curve of $\rm{^{56}Fe}$ over $x$ from $0.05$ to $0.7$ to investigate the order of magnitude effects of $\Delta^-$ term on extracting $\sin^2\theta_W$. It is plotted in Fig.\,\ref{Averageonx}. The $\Delta^-$ term depends on the momentum transfer significantly, which is approximately as same as the NuTeV deviation at $Q^2=20\,\textrm{GeV}^2$. This kind of simple average could overestimate the contributions from large $x$ region, since much of the data in NuTeV came from $x\leq 0.2$ where our assumptions in Eq.\,(\ref{uva1}) and Eq.\,(\ref{dva1}) would not be very suitable.

Since our results shown in Fig.\,\ref{Delta-} and Fig.\,\ref{Averageonx} are mainly derived from parametrizations in EPPS21 and CT18ANLO as well as ``SRC-driven'' nuclear effects assumption, there is little model dependence in our conclusion that the $\Delta^-$ term has a effect of reducing the NuTeV result for $\sin^2\theta_W$. It is important to remember that NuTeV does not measure directly $R_A^-$, but rather measures ratios of experimental candidates within kinematic criteria and compares them to Monte Carlo simulations \cite{NuTeV:2002ryj}. The average $Q^2\approx 20\,\textrm{GeV}^2$ was obtained from Monte Carlo simulation, in fact the actual kinematics of the selected events is poorly known. Therefore our results are not directly applicable to the NuTeV data, but only indicate some general features of ``SRC-driven'' nuclear effects on the extraction of weak-mixing angle.

Previous works indicate the CSV correction may explain roughly half of the NuTeV discrepancy with the SM and the correction from strange quark asymmetry has a significant uncertainty. Other studies of nuclear corrections to the $R_A^-$ include Fermi motion and nuclear shadowing effects \cite{Brodsky:2004qa,Kulagin:2003wz}. Each experiment requires a specific analysis according to the relevant experimental conditions. Principally, all of the corrections mentioned above should be formally incorporated into a reanalysis of the NuTeV data, with good control over various systematic uncertainties, before we could claim that there is no longer any significant discrepancy between the predictions of the SM and the NuTeV data.

The future high luminosity EIC would allow for a series of precision extractions of $\sin^2\theta_W$ \cite{AbdulKhalek:2021gbh}. Besides, the proposed LHC Forward Physics Facility (FPF) is estimated to about $3\%$ precision on measuring $\sin^2\theta_W$ at $Q \!\approx\! 10\,\textrm{GeV}$ \cite{MammenAbraham:2023psg}. Other future experiments include DUNE, Moller, IsoDAR, MESA-P2, etc. \cite{deGouvea:2019wav,SLACE158:2005uay,Alonso:2021kyu,Berger:2015aaa}, which would shed new light on the long-standing NuTeV anomaly.

\section{Summary}
\label{summary}
In summary, we have studied neutral-current neutrino-nucleon DIS with a particular interest in the relationship between SRC and the EMC effect. The ratios of structure function $F_{2(NC)}^{A}(x,Q^2)$ and $x F_{3(NC)}^{A}(x,Q^2)$ are presented to illustrate that the EMC effect in different nuclei can be described by the abundance of SRC pairs and the proposed modification functions in this work are in fact universal.

In addition, we have derived a modified PW relation for nuclei motivated by the correlation between the EMC effect and the SRC scale factor. Taking advantage of this relation, we found the ``SRC-driven'' nuclear effects may account for a substantial fraction of the NuTeV anomaly. This conclusion may have fundamental consequences for our understanding of nucleon structure. Apart from the importance mentioned above, we think the idea of investigating correlation between EMC and SRC effects with modified PW relation $R_A^-$ on the platform of NuTeV and other future neutrino-nucleon scattering experiments in itself is pretty stimulating.

\section*{Acknowledgements}
We thank Prof.\,Wei Wang for the careful reading of the manuscript and suggestions. J.X. is supported in part by National Natural Science Foundation of China under Grant No. 12105247, the China Postdoctoral Science Foundation under Grant No. 2021M702957. X.H.Y. was supported by the Natural Science Foundation of Shandong Province under Grant No. ZR2021QA040. F.H is supported in part by Natural Science Foundation of China under Grant No. 12125503.


\begin{thebibliography}{}
\bibitem{Aubert:1983xm}
J.~J.~Aubert \textit{et al.} [European Muon],
Phys. Lett. B \textbf{123}, 275-278 (1983)
doi:10.1016/0370-2693(83)90437-9
\bibitem{Arneodo:1988aa}
M.~Arneodo \textit{et al.} [European Muon],
Phys. Lett. B \textbf{211}, 493-499 (1988)
doi:10.1016/0370-2693(88)91900-4
\bibitem{Arneodo:1989sy}
M.~Arneodo \textit{et al.} [European Muon],
Nucl. Phys. B \textbf{333}, 1-47 (1990)
doi:10.1016/0550-3213(90)90221-X
\bibitem{Allasia:1990nt}
D.~Allasia \textit{et al.} [New Muon (NMC)],
Phys. Lett. B \textbf{249}, 366-372 (1990)
doi:10.1016/0370-2693(90)91270-L
\bibitem{Gomez:1993ri}
J.~Gomez, R.~G.~Arnold, P.~E.~Bosted, C.~C.~Chang, A.~T.~Katramatou, G.~G.~Petratos, A.~A.~Rahbar, S.~E.~Rock, A.~F.~Sill and Z.~M.~Szalata, \textit{et al.}
Phys. Rev. D \textbf{49}, 4348-4372 (1994)
doi:10.1103/PhysRevD.49.4348.
\bibitem{Seely:2009gt}
J.~Seely, A.~Daniel, D.~Gaskell, J.~Arrington, N.~Fomin, P.~Solvignon, R.~Asaturyan, F.~Benmokhtar, W.~Boeglin and B.~Boillat, \textit{et al.}
Phys. Rev. Lett. \textbf{103}, 202301 (2009)
doi:10.1103/PhysRevLett.103.202301
[arXiv:0904.4448 [nucl-ex]].
\bibitem{Weinstein:2010rt}
L.~B.~Weinstein, E.~Piasetzky, D.~W.~Higinbotham, J.~Gomez, O.~Hen and R.~Shneor,
Phys. Rev. Lett. \textbf{106}, 052301 (2011)
doi:10.1103/PhysRevLett.106.052301
[arXiv:1009.5666 [hep-ph]].
\bibitem{Egiyan:2005hs}
K.~S.~Egiyan \textit{et al.} [CLAS],
Phys. Rev. Lett. \textbf{96}, 082501 (2006)
doi:10.1103/PhysRevLett.96.082501
[arXiv:nucl-ex/0508026 [nucl-ex]].
\bibitem{Hen:2014nza}
O.~Hen, M.~Sargsian, L.~B.~Weinstein, E.~Piasetzky, H.~Hakobyan, D.~W.~Higinbotham, M.~Braverman, W.~K.~Brooks, S.~Gilad and K.~P.~Adhikari, \textit{et al.}
Science \textbf{346}, 614-617 (2014)
doi:10.1126/science.1256785
[arXiv:1412.0138 [nucl-ex]].
\bibitem{Duer:2018sby}
M.~Duer \textit{et al.} [CLAS],
Nature \textbf{560}, no.7720, 617-621 (2018)
doi:10.1038/s41586-018-0400-z
\bibitem{Chen:2016bde}
J.~W.~Chen, W.~Detmold, J.~E.~Lynn and A.~Schwenk,
Phys. Rev. Lett. \textbf{119}, no.26, 262502 (2017)
doi:10.1103/PhysRevLett.119.262502
[arXiv:1607.03065 [hep-ph]].
\bibitem{Xu:2019wso}
J.~Xu and F.~Yuan,
Phys. Lett. B \textbf{801}, 135187 (2020)
doi:10.1016/j.physletb.2019.135187
[arXiv:1908.10413 [hep-ph]].
\bibitem{Hatta:2019ocp}
Y.~Hatta, M.~Strikman, J.~Xu and F.~Yuan,
Phys. Lett. B \textbf{803}, 135321 (2020)
doi:10.1016/j.physletb.2020.135321
[arXiv:1911.11706 [hep-ph]].
\bibitem{Lynn:2019vwp}
J.~E.~Lynn, D.~Lonardoni, J.~Carlson, J.~W.~Chen, W.~Detmold, S.~Gandolfi and A.~Schwenk,
J. Phys. G \textbf{47}, no.4, 045109 (2020)
doi:10.1088/1361-6471/ab6af7
[arXiv:1903.12587 [nucl-th]].
\bibitem{Huang:2021cac}
F.~Huang, J.~Xu and X.~H.~Yang,
Phys. Rev. D \textbf{104}, no.3, 033002 (2021)
doi:10.1103/PhysRevD.104.033002
[arXiv:2103.07873 [hep-ph]].
\bibitem{Hu:2021fxa}
S.~M.~Hu, Y.~S.~Luan and J.~Xu,
[arXiv:2112.14549 [hep-ph]].
\bibitem{Bertone:2018dse}
V.~Bertone, R.~Gauld and J.~Rojo,
JHEP \textbf{01}, 217 (2019)
doi:10.1007/JHEP01(2019)217
[arXiv:1808.02034 [hep-ph]].
\bibitem{Wang:2022kwg}
R.~Wang, N.~N.~Ma and T.~F.~Wang,
Chin. Phys. C \textbf{47}, no.4, 044103 (2023)
doi:10.1088/1674-1137/acb7d0
[arXiv:2207.10980 [nucl-th]].
\bibitem{Hen:2016kwk}
O.~Hen, G.~A.~Miller, E.~Piasetzky and L.~B.~Weinstein,
Rev. Mod. Phys. \textbf{89}, no.4, 045002 (2017)
doi:10.1103/RevModPhys.89.045002
[arXiv:1611.09748 [nucl-ex]].
\bibitem{Eskola:2021nhw}
K.~J.~Eskola, P.~Paakkinen, H.~Paukkunen and C.~A.~Salgado,
Eur. Phys. J. C \textbf{82}, no.5, 413 (2022)
doi:10.1140/epjc/s10052-022-10359-0
[arXiv:2112.12462 [hep-ph]].
\bibitem{Hou:2019efy}
T.~J.~Hou, J.~Gao, T.~J.~Hobbs, K.~Xie, S.~Dulat, M.~Guzzi, J.~Huston, P.~Nadolsky, J.~Pumplin and C.~Schmidt, \textit{et al.}
Phys. Rev. D \textbf{103}, no.1, 014013 (2021)
doi:10.1103/PhysRevD.103.014013
[arXiv:1912.10053 [hep-ph]].
\bibitem{Kumar:2013yoa}
K.~S.~Kumar, S.~Mantry, W.~J.~Marciano and P.~A.~Souder,
Ann. Rev. Nucl. Part. Sci. \textbf{63}, 237-267 (2013)
doi:10.1146/annurev-nucl-102212-170556
[arXiv:1302.6263 [hep-ex]].
\bibitem{Marquet:2019ltn}
C.~Marquet, S.~Y.~Wei and B.~W.~Xiao,
Phys. Lett. B \textbf{802}, 135253 (2020)
doi:10.1016/j.physletb.2020.135253
[arXiv:1909.08572 [hep-ph]].
\bibitem{Wei:2020glg}
S.~y.~Wei,
Phys. Lett. B \textbf{817}, 136356 (2021)
doi:10.1016/j.physletb.2021.136356
[arXiv:2009.06514 [hep-ph]].
\bibitem{NuTeV:2001whx}
G.~P.~Zeller \textit{et al.} [NuTeV],
Phys. Rev. Lett. \textbf{88}, 091802 (2002)
[erratum: Phys. Rev. Lett. \textbf{90}, 239902 (2003)]
doi:10.1103/PhysRevLett.88.091802
[arXiv:hep-ex/0110059 [hep-ex]].
\bibitem{NuTeV:2002ryj}
G.~P.~Zeller \textit{et al.} [NuTeV],
Phys. Rev. D \textbf{65}, 111103 (2002)
[erratum: Phys. Rev. D \textbf{67}, 119902 (2003)]
doi:10.1103/PhysRevD.65.111103
[arXiv:hep-ex/0203004 [hep-ex]].
\cite{Tzanov:2005kr}
\bibitem{Tzanov:2005kr}
M.~Tzanov \textit{et al.} [NuTeV],
Phys. Rev. D \textbf{74}, 012008 (2006)
doi:10.1103/PhysRevD.74.012008
[arXiv:hep-ex/0509010 [hep-ex]].
\bibitem{Schienbein:2007fs}
I.~Schienbein, J.~Y.~Yu, C.~Keppel, J.~G.~Morfin, F.~Olness and J.~F.~Owens,
Phys. Rev. D \textbf{77}, 054013 (2008)
doi:10.1103/PhysRevD.77.054013
[arXiv:0710.4897 [hep-ph]].
\bibitem{Schienbein:2009kk}
I.~Schienbein, J.~Y.~Yu, K.~Kovarik, C.~Keppel, J.~G.~Morfin, F.~Olness and J.~F.~Owens,
Phys. Rev. D \textbf{80}, 094004 (2009)
doi:10.1103/PhysRevD.80.094004
[arXiv:0907.2357 [hep-ph]].
\bibitem{Hirai:2009mq}
M.~Hirai, S.~Kumano and K.~Saito,
AIP Conf. Proc. \textbf{1189}, no.1, 269-275 (2009)
doi:10.1063/1.3274169
[arXiv:0909.2329 [hep-ph]].
\bibitem{Kovarik:2010uv}
K.~Kovarik, I.~Schienbein, F.~I.~Olness, J.~Y.~Yu, C.~Keppel, J.~G.~Morfin, J.~F.~Owens and T.~Stavreva,
Phys. Rev. Lett. \textbf{106}, 122301 (2011)
doi:10.1103/PhysRevLett.106.122301
[arXiv:1012.0286 [hep-ph]].
\bibitem{Paukkunen:2010hb}
H.~Paukkunen and C.~A.~Salgado,
JHEP \textbf{07}, 032 (2010)
doi:10.1007/JHEP07(2010)032
[arXiv:1004.3140 [hep-ph]].
\bibitem{Georgi:1974sy}
H.~Georgi and S.~L.~Glashow,
Phys. Rev. Lett. \textbf{32}, 438-441 (1974)
doi:10.1103/PhysRevLett.32.438
\bibitem{Georgi:1974yf}
H.~Georgi, H.~R.~Quinn and S.~Weinberg,
Phys. Rev. Lett. \textbf{33}, 451-454 (1974)
doi:10.1103/PhysRevLett.33.451
\bibitem{Marciano:1990dp}
W.~J.~Marciano and J.~L.~Rosner,
Phys. Rev. Lett. \textbf{65}, 2963-2966 (1990)
[erratum: Phys. Rev. Lett. \textbf{68}, 898 (1992)]
doi:10.1103/PhysRevLett.65.2963
\bibitem{Davidson:2001ji}
S.~Davidson, S.~Forte, P.~Gambino, N.~Rius and A.~Strumia,
JHEP \textbf{02}, 037 (2002)
doi:10.1088/1126-6708/2002/02/037
[arXiv:hep-ph/0112302 [hep-ph]].
\bibitem{Kurylov:2003by}
A.~Kurylov, M.~J.~Ramsey-Musolf and S.~Su,
Nucl. Phys. B \textbf{667}, 321-348 (2003)
doi:10.1016/S0550-3213(03)00528-5
[arXiv:hep-ph/0301208 [hep-ph]].
\bibitem{Londergan:2003ij}
J.~T.~Londergan and A.~W.~Thomas,
Phys. Rev. D \textbf{67}, 111901 (2003)
doi:10.1103/PhysRevD.67.111901
[arXiv:hep-ph/0303155 [hep-ph]].
\bibitem{Martin:2003sk}
A.~D.~Martin, R.~G.~Roberts, W.~J.~Stirling and R.~S.~Thorne,
Eur. Phys. J. C \textbf{35}, 325-348 (2004)
doi:10.1140/epjc/s2004-01825-2
[arXiv:hep-ph/0308087 [hep-ph]].
\bibitem{Kovalenko:2002xe}
S.~Kovalenko, I.~Schmidt and J.~J.~Yang,
Phys. Lett. B \textbf{546}, 68-77 (2002)
doi:10.1016/S0370-2693(02)02591-1
[arXiv:hep-ph/0207158 [hep-ph]].
\bibitem{Kumano:2002ra}
S.~Kumano,
Phys. Rev. D \textbf{66}, 111301 (2002)
doi:10.1103/PhysRevD.66.111301
[arXiv:hep-ph/0209200 [hep-ph]].
\bibitem{Gluck:2005xh}
M.~Gluck, P.~Jimenez-Delgado and E.~Reya,
Phys. Rev. Lett. \textbf{95}, 022002 (2005)
doi:10.1103/PhysRevLett.95.022002
[arXiv:hep-ph/0503103 [hep-ph]].
\bibitem{Kulagin:2007ju}
S.~A.~Kulagin and R.~Petti,
Phys. Rev. D \textbf{76}, 094023 (2007)
doi:10.1103/PhysRevD.76.094023
[arXiv:hep-ph/0703033 [hep-ph]].
\bibitem{Ball:2009mk}
R.~D.~Ball \textit{et al.} [NNPDF],
Nucl. Phys. B \textbf{823}, 195-233 (2009)
doi:10.1016/j.nuclphysb.2009.08.003
[arXiv:0906.1958 [hep-ph]].
\bibitem{Cloet:2009qs}
I.~C.~Cloet, W.~Bentz and A.~W.~Thomas,
Phys. Rev. Lett. \textbf{102}, 252301 (2009)
doi:10.1103/PhysRevLett.102.252301
[arXiv:0901.3559 [nucl-th]].
\bibitem{Yang:2022xwy}
W.~Yang and X.~Yang,
Phys. Rev. D \textbf{106}, no.9, 093003 (2022)
doi:10.1103/PhysRevD.106.093003
[arXiv:2209.01629 [hep-ph]].
\bibitem{Bentz:2009yy}
W.~Bentz, I.~C.~Cloet, J.~T.~Londergan and A.~W.~Thomas,
Phys. Lett. B \textbf{693}, 462-466 (2010)
doi:10.1016/j.physletb.2010.09.001
[arXiv:0908.3198 [nucl-th]].
\bibitem{Londergan:1998ai}
J.~T.~Londergan and A.~W.~Thomas,
Prog. Part. Nucl. Phys. \textbf{41}, 49-124 (1998)
doi:10.1016/S0146-6410(98)00055-6
[arXiv:hep-ph/9806510 [hep-ph]].
\bibitem{Paschos:1972kj}
E.~A.~Paschos and L.~Wolfenstein,
Phys. Rev. D \textbf{7}, 91-95 (1973)
doi:10.1103/PhysRevD.7.91
\bibitem{Callan:1969uq}
C.~G.~Callan, Jr. and D.~J.~Gross,
Phys. Rev. Lett. \textbf{22}, 156-159 (1969)
doi:10.1103/PhysRevLett.22.156
\bibitem{Frankfurt:1993sp}
L.~L.~Frankfurt, M.~I.~Strikman, D.~B.~Day and M.~Sargsian,
Phys. Rev. C \textbf{48}, 2451-2461 (1993)
doi:10.1103/PhysRevC.48.2451
\bibitem{Segarra:2019gbp}
E.~P.~Segarra, A.~Schmidt, T.~Kutz, D.~W.~Higinbotham, E.~Piasetzky, M.~Strikman, L.~B.~Weinstein and O.~Hen,
Phys. Rev. Lett. \textbf{124}, no.9, 092002 (2020)
doi:10.1103/PhysRevLett.124.092002
[arXiv:1908.02223 [nucl-th]].
\bibitem{Alde:1990im}
D.~M.~Alde, H.~W.~Baer, T.~A.~Carey, G.~T.~Garvey, A.~Klein, C.~Lee, M.~J.~Leitch, J.~W.~Lillberg, P.~L.~McGaughey and C.~S.~Mishra, \textit{et al.}
Phys. Rev. Lett. \textbf{64}, 2479-2482 (1990)
doi:10.1103/PhysRevLett.64.2479
\bibitem{Bertsch:1993vx}
G.~F.~Bertsch, L.~Frankfurt and M.~Strikman,
Science \textbf{259}, 773-774 (1993)
doi:10.1126/science.259.5096.773
\bibitem{Fomin:2011ng}
N.~Fomin, J.~Arrington, R.~Asaturyan, F.~Benmokhtar, W.~Boeglin, P.~Bosted, A.~Bruell, M.~H.~S.~Bukhari, M.~E.~Christy and E.~Chudakov, \textit{et al.}
Phys. Rev. Lett. \textbf{108}, 092502 (2012)
doi:10.1103/PhysRevLett.108.092502
[arXiv:1107.3583 [nucl-ex]].
\bibitem{Arrington:2012ax}
J.~Arrington, A.~Daniel, D.~Day, N.~Fomin, D.~Gaskell and P.~Solvignon,
Phys. Rev. C \textbf{86}, 065204 (2012)
doi:10.1103/PhysRevC.86.065204
[arXiv:1206.6343 [nucl-ex]].
\bibitem{Kulagin:2003wz}
S.~A.~Kulagin,
Phys. Rev. D \textbf{67}, 091301 (2003)
doi:10.1103/PhysRevD.67.091301
[arXiv:hep-ph/0301045 [hep-ph]].
\bibitem{Brodsky:2004qa}
S.~J.~Brodsky, I.~Schmidt and J.~J.~Yang,
Phys. Rev. D \textbf{70}, 116003 (2004)
doi:10.1103/PhysRevD.70.116003
[arXiv:hep-ph/0409279 [hep-ph]].
\bibitem{AbdulKhalek:2021gbh}
R.~Abdul Khalek, A.~Accardi, J.~Adam, D.~Adamiak, W.~Akers, M.~Albaladejo, A.~Al-bataineh, M.~G.~Alexeev, F.~Ameli and P.~Antonioli, \textit{et al.}
Nucl. Phys. A \textbf{1026}, 122447 (2022)
doi:10.1016/j.nuclphysa.2022.122447
[arXiv:2103.05419 [physics.ins-det]].
\bibitem{MammenAbraham:2023psg}
R.~Mammen Abraham, S.~Foroughi-Abari, F.~Kling and Y.~D.~Tsai,
[arXiv:2301.10254 [hep-ph]].
\bibitem{deGouvea:2019wav}
A.~de Gouvea, P.~A.~N.~Machado, Y.~F.~Perez-Gonzalez and Z.~Tabrizi,
Phys. Rev. Lett. \textbf{125}, no.5, 051803 (2020)
doi:10.1103/PhysRevLett.125.051803
[arXiv:1912.06658 [hep-ph]].
\bibitem{SLACE158:2005uay}
P.~L.~Anthony \textit{et al.} [SLAC E158],
Phys. Rev. Lett. \textbf{95}, 081601 (2005)
doi:10.1103/PhysRevLett.95.081601
[arXiv:hep-ex/0504049 [hep-ex]].
\bibitem{Alonso:2021kyu}
J.~Alonso, C.~A.~Arg\"uelles, A.~Bungau, J.~M.~Conrad, B.~Dutta, Y.~D.~Kim, E.~Marzec, D.~Mishins, S.~H.~Seo and M.~Shaevitz, \textit{et al.}
Phys. Rev. D \textbf{105}, no.5, 052009 (2022)
doi:10.1103/PhysRevD.105.052009
[arXiv:2111.09480 [hep-ex]].
\bibitem{Berger:2015aaa}
N.~Berger, K.~Aulenbacher, S.~Baunack, D.~Becker, J.~Diefenbach, M.~Gericke, K.~Gerz, R.~Herbertz, K.~Kumar and F.~Maas, \textit{et al.}
J. Univ. Sci. Tech. China \textbf{46}, no.6, 481-487 (2016)
doi:10.3969/j.issn.0253-2778.2016.06.006
[arXiv:1511.03934 [physics.ins-det]].



\end{thebibliography}
\end{document}